\documentstyle[12pt] {article}
\begin {document}
\parindent=15pt
\begin{center}
\vskip 1.5 truecm
{\bf HOW TO ACCOUNT FOR THE INTERFERENCE CONTRIBUTIONS IN MONTE CARLO
SIMULATIONS}\\
\vspace{.5cm}
Yu.M.Shabelski \\
\vspace{.5cm}
Petersburg Nuclear Physics Institute, \\
Gatchina, St.Petersburg 188350 Russia \\
\end{center}
\vspace{1cm}
\begin{abstract}
The diagram technique allows one to calculate the correction factors
which can be used in Monte Carlo simulation of some processes. This is
equivalent to the calculation with accounting for all or some part of
the interference contributions. The example is presented for the
simplest case of inelastic deuteron-deuteron interactions.
\end{abstract}
\vspace{3cm}

E-mail: $\;$ shabel@vxdesy.desy.de \\

\newpage


We will consider the discussed problem for the concrete case of
intranuclear cascade model with Monte Carlo simulation of events. This
model is rather popular until now, especially at not very high energies
\cite{GT,AGT}. It is well-known that all interference contributions are
lost in such simulation because in the Monte Carlo method we can add
probabilities but not amplitudes. In many cases it leads to not so large
errors because the interference contributions are not dominated. However
until now it was not possible even to estimate their role qualitatively.

In the present paper we will show that it is possible, as a minimum
in some special cases, to calculate the correction factors. Use of them
is equivalent to the account, as a minimum, some part of interference
contributions.

Let us consider for simplicity the deuteron-deuteron interaction at
energy a little smaller than 1 GeV per nucleon. In this situation
only one secondary pion can be produced in each nucleon-nucleon
interaction, and a secondary nucleon after the first inelastic collision
practically can not produce another pion because it has no enough
energy. So the main source of pion production in the considered case
is the process of one-nucleon pair inelastic interaction, Fig. 1a.
Double-nucleon pair interaction has smaller probability, the
process of Fig. 1b gives some correction to the cross section of one
pion production whereas the process of Fig. 1c is qualitatively
different because it leads to the two pion production in one event.

There exist also a lot of processes with elastic rescattering of
secondary nucleons and pions but they can not change the pion
multiplicity (except of the case of pion absorption by secondary
nucleon).

Let us consider now the processes of Fig. 1 from the point of view
of unitarity condition. The modulo squared amplitude of Fig. 1a is
shown as a cut of elastic $dd$ scattering amplitude in Fig. 2a. If
the cross section determined by the imaginary part of the amplitude
Fig. 2a is equal to $\Delta_1$, the cross section of the process
Fig. 1a is equal to
\begin{equation}
\sigma_{1a} = \Delta_1 \frac{\sigma_{NN}^{inel}}{\sigma_{NN}^{tot}}
\end{equation}
(we neglect the difference in $pp$, $pn$ and $nn$ cross sections
for simplicity).

The modulo squared amplitudes of Fig. 1b and 1c correspond to the cuts
of another diagram of $dd$ elastic scattering amplitude which are shown
in Fig. 2b and 2c. The only difference between them is that in the case
of Fig. 1b we should take one cutted nucleon-nucleon blob with inelastic
intermediate state and another one with elastic $NN$ scattering, whereas
in the case of Fig. 1c the inelastic intermediate states in both cutted
blobs should be taken. So if the contribution of the diagram Fig. 2b (2c)
to the total $dd$ cross section is equal to $\Delta_2$, the cross
sections of the processes Fig. 1b and 1c are
\begin{equation}
\sigma_{1b} = 4 \Delta_2 \frac{\sigma_{NN}^{inel}}{\sigma_{NN}^{tot}}
\frac{\sigma_{NN}^{el}}{\sigma_{NN}^{tot}}
\end{equation}
and
\begin{equation}
\sigma_{1c} = 2 \Delta_2 \frac{\sigma_{NN}^{inel}}{\sigma_{NN}^{tot}}
\frac{\sigma_{NN}^{inel}}{\sigma_{NN}^{tot}} \;,
\end{equation}
respectively. Factor two in both these Eqs. come from the AGK cutting
rules \cite{AGK,Sh} and another factor two in Eq. (2) comes from
combinatoric.

The diagram Fig. 1c can not interfere with another diagrams of
Fig. 1 because it contain two pions in the final state. However the
diagrams of Fig. 1a and 1b can interfere that corresponds to the
intermediate state of elastic $dd$ amplitude shown in Fig. 2c. In
accordance with AGK cutting rules this contribution to cross
section is
\begin{equation}
\sigma_{1a,1b} = -4 \Delta_2 \frac{\sigma_{NN}^{inel}}{\sigma_{NN}^{tot}}
\end{equation}
This cross section can be calculated inside the Monte Carlo code via
the value of $\sigma_{1c}$ (or the correspondent number of events) :
\begin{equation}
\sigma_{1a,1b} = -2 \sigma_{1c}
\frac{\sigma_{NN}^{tot}}{\sigma_{NN}^{inel}}
\end{equation}

So we should multiply all distributions, histograms,
multiplicities, etc., coming from the sum of events from Fig. 1a and
Fig. 1b processes by the factor
\begin{equation}
R = \frac {\sigma_{1a} + \sigma_{1b} - \sigma_{1a,1b}}
{\sigma_{1a} + \sigma_{1b}} < 1
\end{equation}
and only after that add the events from the processes of Fig. 1c. In
particular one can see that the mean multiplicity of produced pions will
increase because we add smaller number of one-pion events with the same
number of two-pion events.


The similar calculations can be fulfilled with the help of the same AGK
cutting rules for more realistic cases of hadron-nucleus and
nucleus-nucleus interactions and possibly in some another cases.
Of course there exist many another interference contributions
connected, say, with final state interactions, etc. which will be
not accounted by the similar way. However sometimes these contributions
can be not essential. So one can see that the combination of Monte Carlo
code which allow one to calculate, say, some angular distributions of
produced pions, and AGK cutting rules gives the possibility to increase
the accuracy of calculations. Possibly the similar approach can be used
in another cases where Monte Carlo simulations are used.

I am grateful to A.Capella for useful discussions.

This work is supported by INTAS grant 93-0079.


\begin{center}
{\bf Figure captions}\\
\end{center}

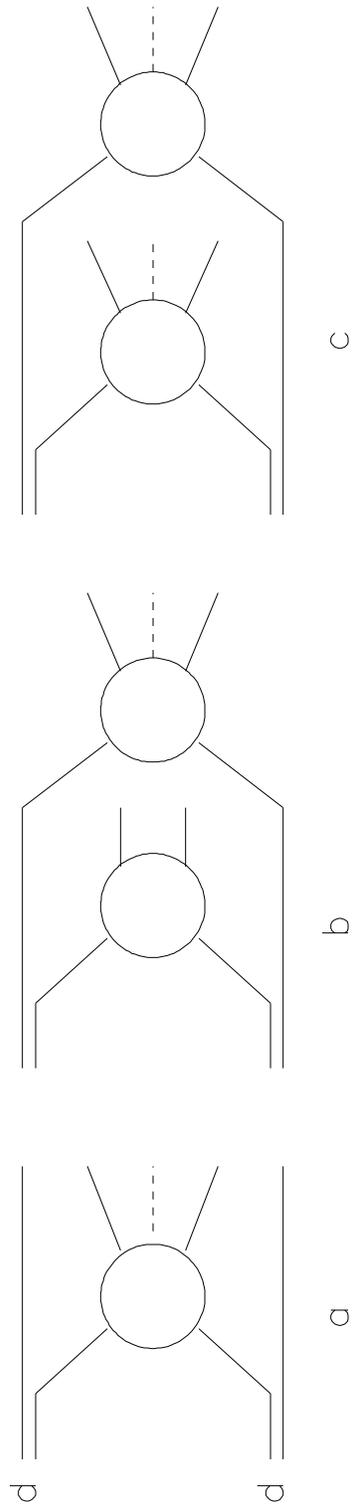
Fig. 1. Diagrams for pion production in not high energy
deuteron-deuteron interactions.

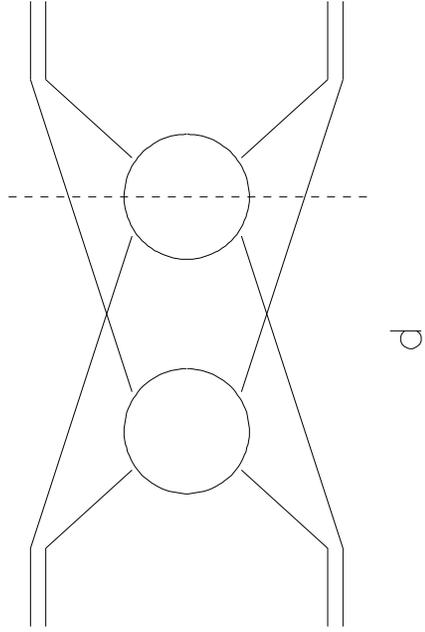
Fig. 2. The intermediate states of elastic $dd$ amplitude which
correspond (a, b and c) to the modulo squares of the diagrams of Fig. 1
and (d) to the interference of amplitudes Fig. 1a and 1b.


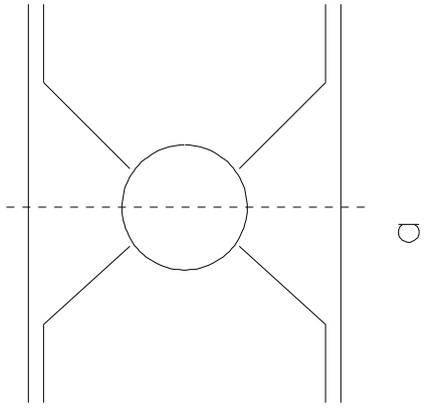
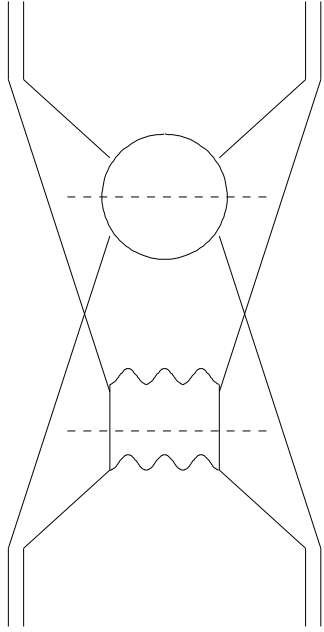
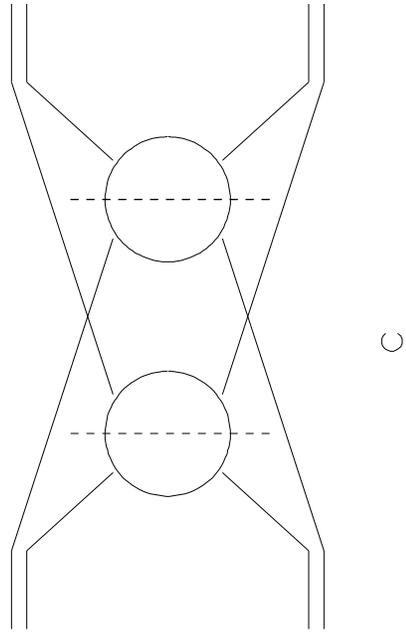
\end{document}